\documentclass[12pt]{iopart}
\pdfoutput=1
\usepackage{iopams}  

\expandafter\let\csname equation*\endcsname\relax
\expandafter\let\csname endequation*\endcsname\relax
\usepackage{mathtools} 

\usepackage{graphicx,graphics,epsfig,subfigure,times,bm,bbm,amssymb,amsmath,amsthm,mathrsfs,MnSymbol}
\usepackage{gensymb}
\usepackage{amsfonts}
\usepackage{float}
\usepackage[matrix,frame,arrow]{xypic}
\usepackage[pdfstartview=FitH]{hyperref}

\usepackage[pdftex]{color}
\usepackage{hypernat}
\usepackage{braket}
\usepackage{enumerate}
\usepackage[normalem]{ulem}
\usepackage[usenames,dvipsnames]{xcolor}
\usepackage{multirow}
\usepackage{mathtools}
\usepackage{bbm}
\usepackage{titletoc}

\usepackage{cite}

\definecolor{orange}{rgb}{1,0.5,0}

\newcommand{\RNum}[1]{\uppercase\expandafter{\romannumeral #1\relax}}

\newcommand{\ignore}[1]{}
\usepackage{geometry}

\ignore{
	\documentclass[eprintnumbers,amsmath,amssymb,onecolumn,a4paper,caption ]{article}
	\usepackage{amsfonts}
	\usepackage{amssymb}
	\usepackage{mathrsfs}
	\usepackage{bbm}
	\usepackage{mathrsfs}
	\usepackage{dcolumn}
	\usepackage{bm}
	\usepackage{times,epsfig,amssymb,amsmath}
	\usepackage{float}
	\usepackage{subfigure}
	\usepackage{geometry}\geometry{left=2.5cm,right=2.5cm,top=3cm,bottom=3cm}

	\usepackage{color}

}

\hypersetup{
	colorlinks=true,	
	linkcolor=red,     	
	citecolor=blue,  	
	filecolor=magenta, 	
	urlcolor=blue,     	
	runcolor=cyan
}
\begin{document}

\title{Multipartite Entanglement in Crossing the Quantum Critical Point}

\author{Hao-Yu~Sun}
\address{Institute of Physics, Chinese Academy of Sciences, Beijing 100190, China}
\address{School of Physical Sciences, University of Chinese Academy of Sciences, Beijing 100190, China}

\author{Zi-Yong~Ge}
\address{Theoretical Quantum Physics Laboratory, Cluster for Pioneering Research, RIKEN, Wako-shi, Saitama 351-0198, Japan}

\author{Heng~Fan\thanks{Corresponding author: \mailto{hfan@iphy.ac.cn}}}
\ead{\mailto{hfan@iphy.ac.cn}}
\address{Institute of Physics, Chinese Academy of Sciences, Beijing 100190, China}	
\address{School of Physical Sciences, University of Chinese Academy of Sciences, Beijing 100190, China}
\address{Songshan Lake Materials Laboratory, Dongguan 523808, Guangdong, China}
\address{CAS Center for Excellence in Topological Quantum Computation, UCAS, Beijing 100190, China}
\address{Beijing Academy of Quantum Information Sciences, Beijing 100193, China}


\begin{abstract}
We investigate the multipartite entanglement for a slow quantum quench crossing a critical point.
We consider the quantum Ising model and the Lipkin-Meshkov-Glick model, 
which are local and full-connected quantum systems, respectively. 
The multipartite entanglement is quantified by quantum Fisher information with the generator
defined as the operator of the ferromagnetic order parameter.
The quench dynamics begins with a ground state in a paramagnetic phase,
and then the transverse field is driven slowly to cross a quantum critical point, and ends with a zero transverse field.
For the quantum Ising model, based on methods of matrix product states, we calculate the quantum Fisher information density of the final state. 
Numerical results of both linear and nonlinear quenches show that the quantum Fisher information density of the final state scales as a power law of the quench rate, 
which overall conforms to the prediction of the Kibble-Zurek mechanism with a small correction. 
We show that this correction results from the long-range behaviors.
We also calculate the quantum Fisher information density in the Lipkin-Meshkov-Glick model. 
The results show that the scaling of quantum Fisher information in this full-connected system conforms to the Kibble-Zurek mechanism better,
since the long-range physics cannot be defined in this nonlocal system. 
Our results reveal that the multipartite entanglement provides an alternative viewpoint to understand the dynamics of quantum phase transitions,
specifically, the nontrivial long-range physics. 
\end{abstract}
%
%

%
\maketitle
%
%

\section{Introduction}\label{introduction}

Quantum entanglement plays a crucial role in quantum information processing and quantum computations.
For instance, the bipartite entanglement, generally represented by von Neumann or R\'enyi entropies, is an essential ingredient for supporting quantum advantages in quantum computations.
The multipartite entanglement, quantified by spin squeezing~\cite{Hyllus2012, Toth2012, Hauke2016} or quantum Fisher information (QFI)~\cite{Zhang2018, Pezze2017, Wu2005}, 
is a type of indispensable quantum resource to perform quantum metrology beyond the standard quantum limit~\cite{Braunstein1994, Giovannetti2006, Pezze2016, Giovannetti2011, Peng2020, Wang2021}.
In addition, quantum entanglement can also be applied to understand quantum many-body physics~\cite{Amico2008, Sachdev2009, Suzuki2013, Zeng2019} both in theory~\cite{Fromholz2020} and experiment~\cite{Xia2009, Xu2018}.
The bipartite entanglement can be used to identify different quantum phases and critical phenomena~\cite{Kitaev2006, Levin2006}.
For example, there exists a universal subleading entanglement entropy, termed topological entanglement entropy~\cite{Kitaev2006, Levin2006, Li2008, Eisert2010}, in intrinsic topological order systems.
In critical systems, conformal field theories (CFT) predict that the entanglement entropy is logarithmically divergent with respect to the system size, where the factor is proportional to the central charge~\cite{Vidal2003, Holzhey1994, Cardy2016}.
Recently, the multipartite entanglement is also applied to witness the symmetry-protected topological order ~\cite{Osterloh2002, Guehne2005, Hofmann2014, Movassagh2016}.

The Kibble-Zurek mechanism~(KZM) describes the universal dynamics of the second-order phase transitions (QPTs). 
It was first proposed to understand thermodynamic phase transitions in cosmological physics~\cite{Kibble1976, Kibble1980}, 
which had been introduced to classical second-order phase transitions~\cite{Zurek1985, Zurek1996}, 
and verified by numerical simulations including the density of kinks after a quench~\cite{Laguna1997}, 
vortex in two dimensions~\cite{Yates1998} and inhomogeneous quenches~\cite{Dziarmaga1999}.
In addition, KZM is also valid in quantum phase transitions~\cite{Polkovnikov2005, Zurek2005, Dziarmaga2005}, 
encompassing symmetry-breaking phase transitions~\cite{GomezRuiz2019, GomezRuiz2020}, 
topological phase transitions~\cite{Lee2015, Liou2018}, 
localization-delocalization transitions~\cite{Roosz2014}, 
and deconfined quantum criticality~\cite{Huang2020}. 
Meanwhile, KZM can be applied to understand adiabatic dynamics~\cite{Barankov2008}
and quantum annealers~\cite{Bando2020}.
Recently, it is also studied in other aspects of quantum many-body physics~\cite{Kou2023, Suzuki2024, Jamadagni2024}.
Moreover, KZM has been demonstrated experimentally in a variety of systems, such as 
solid materials~\cite{Chuang1991, Du2023}, 
cold atoms~\cite{Navon2015, Ko2019},
Rydberg simulators~\cite{Keesling2019}, 
superconducting circuits~\cite{Monaco2006}, 
and trap-ions~\cite{Cui2020}.

At the critical point, the correlation length and the relaxation time diverge, which is known as critical slowing down.
Thus, for a slow quench dynamics crossing a critical point, the adiabatic condition will break down,  and the system can excite topological defects in the final state~\cite{Dziarmaga2006}.
KZM predicts that the density of topological defects $n_d$ has a universal power-law scaling with respect to the quench rate $\tau_Q$: $n_d=\tau_Q^{-d\nu/(z\nu+1)}$~\cite{Kibble1976}, where $d$ is the spatial dimension, $\nu$ and $z$ are the correlation length and dynamical critical exponent, respectively.

In addition to the topological defects, KZM can also be used to predict other physical quantities~\cite{Polkovnikov2011}.
It is shown that the diagonal entropy after a slow quantum quench crossing the critical point also satisfies the Kibble-Zurek scaling~\cite{Grandi2010, Polkovnikov2011a, Mukherjee2008, Grandi2010a}.
For the bipartite entanglement, the von Neumann entropy is found to scale logarithmically with respect to the quench rate, which is closely related to KZM~\cite{Cincio2007, Pollmann2010}.

The multipartite entanglement is more complicated, which has not been depicted uniquely and clearly. 
Here, the density of topological defects (e.g., domain walls) is a local observable, which generally relates to the short-range correlation function. 
Thus, topological defects cannot describe the long-range physics.
In contrast, the multipartite entanglement is always a nonlocal observable. 
Therefore, one open question may arise:  whether there exists a relation between multipartite entanglement and KZM in crossing a quantum critical point.

In this paper, we focus on the multipartite entanglement after a slow quench crossing a quantum critical point. 
We discuss the quantum Ising model and the Lipkin-Meshkov-Glick~(LMG) model, 
which are local and full-connected quantum systems, respectively. 
The multipartite entanglement is witnessed by the QFI density, where the generator is chosen from the symmetries that relate to the phase transitions. 
For the quantum Ising model, we calculate the QFI density of the final state, based on the matrix product state~(MPS) method. 
We find that the QFI density of the final state exhibits a power-law scaling with respect to the quench rate for both linear and nonlinear quenches.
The power-law exponent is close to the KZM scaling with a small correction. 
We show that this correction is from the long-range physics of the final states, which is beyond KZM.
For the LMG model, we calculate the complement of the QFI density.
The result conforms to the prediction of KZM, where the correction can be nearly neglected. 
In addition, based on the above results, we present a phenomenological description of critical slowing down in QPTs from the perspective of multipartite entanglement. 

This paper is organized as follows:
In Sec.~\ref{kzm}, we review KZM in quantum phase transitions for linear and nonlinear cases. 
The multipartite entanglement and QFI are introduced in Sec.~\ref{qfi_me}.
In Sec.~\ref{ising}, we introduce the quantum Ising model including the spectrum, 
and the QFI density of the final states after both linear and nonlinear quantum quenches. 
In Sec.~\ref{lmg}, we study QFI of the quenched state in the LMG model.  
In Sec.~\ref{discussion}, we present a phenomenological discussion of our results. 
Finally, in Sec.~\ref{summary}, we summarize our results and present an outlook.

\section{Kibble-Zurek Mechanism in Quantum Phase Transition}\label{kzm}
In this section, we briefly review KZM in quantum phase transitions for both linear and nonlinear quenches~\cite{Suzuki2013, Polkovnikov2005, Dziarmaga2010}. 
We consider a quantum system driven by a parameter $g$, and the critical point is at $g=g_c$. 
Near the critical point, the correlation length $\xi$ and the energy gap $\Delta$ closely relate to the critical exponents $\nu$ and $z$, i.e.,
\begin{eqnarray}\label{cexp}
\begin{aligned}
&\xi \sim |g-g_c|^{-\nu},  \\
&\Delta \sim |g-g_c|^{z\nu}.
\end{aligned}
\end{eqnarray}
Now we consider a slow quench, where the parameter $g$ becomes a time-dependent function $g(t)$. 
We first consider a linear quench, where we have
\begin{eqnarray}\label{linearized}
g(t)=-\frac t {\tau_Q} + g_c.
\end{eqnarray}
When $t=0$, the system approaches the critical point $g(0)=g_c$. 
We note that other functions of $g(t)$ can always be linearized near the critical point once $g'(t=0)\neq 0$.

Initially, i.e., $t\rightarrow -\infty$, the system starts with the corresponding ground state, where the energy gap $\Delta$ is large enough.
Thus, the adiabatic condition is satisfied, and the system keeps in the ground state. 
With the increase of $t$, the adiabatic condition will be violated near $t=0$, i.e., in the vicinity of the critical point.
The corresponding time range for violating the adiabatic condition is labeled by $[-\hat{t},\hat{t}]$,
and excitations can be excited in this regime.
When $t>\hat t$, the adiabatic condition will recover.

Now we calculate the scale of $\hat{t}$. 
In the quantum phase transition, the coherent time $\tau$ is determined by the inverse of the energy gap
\begin{eqnarray}
\tau\sim\Delta^{-1}. 
\end{eqnarray}
From Eq.~(\ref{cexp}), we can find that the energy gap scales as 
\begin{eqnarray}\label{gap}
\Delta\sim|g-g_c|^{z\nu}=|t/\tau_Q|^{z\nu}.
\end{eqnarray}
When $|t|=\hat{t}$, the coherent time satisfies 
\begin{eqnarray}
\tau \sim |\dot{\Delta}/{\Delta}|^{-1}.
\end{eqnarray}
Thus, we have 
\begin{eqnarray}
\hat{t}\sim\tau_Q^{z\nu/z\nu+1},
\end{eqnarray}
and the corresponding energy gap scales as
\begin{eqnarray}
\hat{\Delta}\sim \tau_Q^{-z\nu/z\nu+1}.
\end{eqnarray}
Therefore, the excitation density scales
\begin{eqnarray}\label{linear_density}
n_d\sim \hat{\Delta}^{d/z}\sim\tau_Q^{-d\nu/z\nu+1},
\end{eqnarray}
where $d$ is the system dimension.
The above picture is known as KZM.

Now we also consider a nonlinear slow quench, where we have $g'(t=0)=0$.
Without loss of generality, we can use the following geometric function to describe the nonlinear slow quench, i.e.,
\begin{eqnarray}
g(t)=-\text{sgn}(t)\left| \frac t{\tau_Q}\right|^\alpha+g_c,
\end{eqnarray}
where $\text{sgn}(t)$ is a sign function.
Here, we discuss KZM of a nonlinear quench from the viewpoint of parameter-dependent critical exponents.
Now, we assume $g$ is a function of another parameter $\lambda$, i.e., $g = g(\lambda)$.
Let $g_c = g(\lambda_c)$, and without loss of generality, we fix $\lambda_c=0$.
Thus, Eq.~(\ref{cexp}) can be rewritten as
\begin{eqnarray}\label{lam}
\begin{aligned}
&\xi\sim|g(\lambda)-g(\lambda_c)|^{-\nu}\sim |\lambda g'(\lambda)|^{-\nu}, \\
&\Delta\sim|g(\lambda)-g(\lambda_c)|^{z\nu}\sim |\lambda g'(\lambda)|^{z\nu}.
\end{aligned}
\end{eqnarray}
In addition, if we consider $\lambda$ as the driven parameter of QPT, where $\lambda_c$ is the critical point,
then the corresponding critical exponents, labeled by $\nu_\lambda$ and $z_\lambda$,  can be defined by
\begin{eqnarray}\label{cexp_lam}
\begin{aligned}
&\xi\sim\lambda^{-\nu_\lambda}, \\
&\Delta\sim \lambda^{z_\lambda\nu_\lambda}.
\end{aligned}
\end{eqnarray}

According to Eqs.~(\ref{cexp}, \ref{cexp_lam}), when $g'(\lambda_c) \neq 0$,
we can find that $\nu_\lambda=\nu$ and $z_\lambda=z$.
However, if $g'(\lambda_c) = 0$, the situation becomes different.
If we let $\lambda = (g-g_c)^{1/\alpha}$, then $g(\lambda) = \lambda^\alpha + g_c$.
Thus, in the vicinity of $\lambda_c$, we can calculate the derivative of $g(\lambda) $ as $g'(\lambda)=\alpha\lambda^{\alpha-1}$.
Therefore, from Eqs.~(\ref{cexp}, \ref{lam}), we can obtain the  critical exponents  with respective to $\lambda$ as
\begin{eqnarray}\label{lam2}
\begin{aligned}
&\nu_\lambda = \alpha\nu, \\
&z_\lambda\ = z.
\end{aligned}
\end{eqnarray}
In addition, we have $\lambda(t) = t/\tau_Q$.
Thus, this nonlinear quench with respective to $g$ is transformed to a linear quench for $\lambda$,
where the results of linear quench can be applied directly. 
For instance, we can use KZM to obtain the topological defect density of the final state as
\begin{eqnarray}
n_d\sim \tau_Q^{- d \nu_\lambda/(z \nu_\lambda +1)}=\tau_Q^{-\alpha d \nu/\alpha z \nu + 1},
\end{eqnarray}
which is consistent with the results in Refs.~\cite{Sen2008, Cincio2007}.

\section{Quantum Fisher Information and Multipartite Entanglement}\label{qfi_me}

Generally, for an operator $\hat O$ and a quantum state with density matrix $\hat \rho=\sum_ip_i\ket{\psi_i}\bra{\psi_i}$, where $\langle\psi_i|\psi_j\rangle=\delta_{ij}$, the corresponding QFI can be defined as
\begin{eqnarray}
F_Q :=2\sum_{p_i+p_j\neq 0}\frac{(p_i-p_j)^2|\bra{\psi_i}\hat{O}\ket{\psi_j}|^2}{p_i+p_j}.
\end{eqnarray}
For a pure state $\hat \rho=\ket{\psi}\bra{\psi}$,  QFI can be simplified as
\begin{eqnarray}\label{qfi}
F_Q=4(\Delta \hat{O})^2=4(\bra{\psi} \hat{O}^2\ket{\psi}-\bra{\psi} \hat{O}\ket{\psi}^2).
\end{eqnarray}
In addition, we can define the QFI density of a system with $N$ particles as
\begin{eqnarray}
f_Q :=\frac 1 N F_Q.
\end{eqnarray}
{Here, to understand the connection between the QFI density and the multipartite entanglement, 
we introduce the concepts of $k$-producible and $k$-entangled states. 
If a state is $k$-producible, it can be decomposed as a product of states of a series of subsystems, 
where each subsystem contains no more than $k$ particles. 
In addition, a state is $k$-entangled, if it is $k$-producible but not $(k-1)$-producible, 
i.e., it contains at least one decomposed state with $k$ particles entangled. 
For a $k$-producibel state, QFI has an upper bound corresponding to the $k$ entangled particles~\cite{Hyllus2012}
\begin{eqnarray}\label{fq_bound}
F_Q \leq \big[\frac{N}{k}\big]k^2+(N-\big[\frac{N}{k}\big]k)^2,
\end{eqnarray}
where $[\frac{N}{k}]$ represents the maximum integer that is smaller than $\frac{N}{k}$.
Therefore, if the QFI density $f_Q$ of a $k$-producible state violates bound, 
the entangled particles of the state must be larger than $k$, i.e., at least $k+1$. 
Thus, according to Eq.~(\ref{fq_bound}), 
the physical significance of the QFI density is that there are at least $[f_Q]+1$ particles entangled.
}

Thus, QFI is one of the quantifications of the multipartite entanglement.
Generally, QFI is an important quantum resource in quantum metrology,
which quantifies the maximal precision of the parameter estimation.
Recently, the multipartite entanglement is also applied to understand quantum many-body physics, e.g., as a witness of topological phases~\cite{Osterloh2002, Movassagh2016}.
Naturally, one may wonder whether the scaling behaviors of the multipartite entanglement can also be described as KZM.
Next, we will study QFI of the final state quenched from the initial ground state crossing the critical point with the rate $\tau_Q$.

\section{Quantum Ising Model}\label{ising}
\subsection{KZM in Quantum Ising Model}\label{kzm_ising}
Here, we consider a one-dimensional spin$-\frac12$ quantum Ising model with transverse fields, of which Hamiltonian reads
\begin{eqnarray}\label{model:hamiltonian_spin}
H=-J\sum_{n}^{N}\hat{\sigma}_n^z\hat{\sigma}_{n+1}^z - g\sum_{n}^{N}\hat{\sigma}_n^x,
\end{eqnarray}
where $N$ is the system size, $\hat{\sigma}^{x/z}$ is the Pauli matrix,
$J>0$ is the nearest-neighbor interaction strength and is fixed to 1 in the following discussions,
and $g$ is the strength of the transverse field. 
By the Jordan-Wigner transformation, the Hamiltonian~(\ref{model:hamiltonian_spin}) can be written as a free spinless fermion chain:
\begin{eqnarray}\label{model:hamiltonian_fermion}
H=-\sum_{n}\Big(\hat{c}_n^\dagger \hat{c}_{n+1}+\hat{c}_n \hat{c}_{n+1}+h.c.\Big)+\frac g2 \sum_{n}\Big(1-2\hat{c}_n^\dagger \hat{c}_n\Big),
\end{eqnarray}
where $\hat{c}_n$ ($\hat{c}_n^\dagger$) is the fermionic annihilate (create) operator at site $n$.
If we choose periodic boundary conditions for the spin system, i.e., $\hat{\sigma}_{N+1}=\hat{\sigma}_1$,
then it corresponds to a periodic/anti-periodic boundary condition in the fermionic system for the odd/even fermion parity.
Now we introduce a Fourier transform,
\begin{eqnarray}
\hat{c}_n=\frac{1}{\sqrt{N}}\sum_k \hat{c}_ke^{ikn},
\end{eqnarray}
where
\begin{eqnarray}
k=\pm\frac12 \frac {2\pi}{N},\cdots,\pm\frac{N-3}{2}\frac {2\pi}{N}, \pm\frac{N-1}{2}\frac {2\pi}{N}.
\end{eqnarray}
Thus, the Hamiltonian~(\ref{model:hamiltonian_fermion}) can be rewritten in the momentum space as (where the constant term is neglected)
\begin{eqnarray}\label{model:hamiltonian_k}
H=-\sum_k\Big(e^{ik}\hat{c}_k^\dagger \hat{c}_k +e^{-ik}\hat{c}_k\hat{c}_{-k}+\text{h.c.} \Big)- g \sum_k \hat{c}_k^\dagger \hat{c}_k. \notag \\
\end{eqnarray}
With Bogoliubov transformation
\begin{eqnarray}
\hat{c}_k=u_k\hat{\gamma}_k+v_k \hat{\gamma}_{-k}^\dagger,
\end{eqnarray}
the Hamiltonian~(\ref{model:hamiltonian_k}) can be diagonalized as
\begin{eqnarray}
\begin{pmatrix}
g-\cos k	&	\sin k\\
\sin k 	&	-g+\cos k
\end{pmatrix}
\begin{pmatrix}u_k\\v_k\end{pmatrix}=
\varepsilon_k \begin{pmatrix}u_k\\v_k\end{pmatrix}.
\end{eqnarray}
The mode $(u_k, v_k)^\mathrm{T}$ is the eigenstate of the Bogoliubov-de Gennes (B-dG) equation with eigenvalues
\begin{eqnarray}
\varepsilon_k=\pm\sqrt{(g-\cos k)^2+\sin^2 k}.
\end{eqnarray}
When $|g| = g_c=1$, the system is gapless, i.e., at the critical point.
When $|g| > g_c$, the spins tend to polarize in $x$-direction at the ground state, so the system is in a paramagnetic phase.
When $|g| < g_c$, the system emerges a long-range order, i.e., in a ferromagnetic phase.

Now, we consider a slow quench dynamics of the quantum Ising model.
Here, the transverse field is time-dependent and satisfies
\begin{eqnarray}
g(t)=-\frac t{\tau_Q}+1,
\end{eqnarray}
where $ -\infty\leq t \leq \tau_Q$.
The initial state is the ground state for $g = \infty$, i.e. all spins are polarized to the ground state of $-\hat{\sigma}_n^x$.
During the slow quench to a zero transverse field, the system needs to cross the critical point, leading to a breaking of adiabaticity.
We label the instantaneous eigenstate at time $t$ as $\ket{\psi(t)}$.
Thus, the final state can be denoted by $\ket{\psi(\tau_Q)}$, which satisfies $\tilde{\gamma}_k\ket{\psi(\tau_Q)}=0$.
Here, $\tilde{\gamma}_k$ is defined by the time-dependent Bogoliubov transformation
\begin{eqnarray}
\hat{c}_k=u_k(t)\tilde{\gamma}_k+v_k(t) \tilde{\gamma}_{-k}^\dagger,
\end{eqnarray}
where $[u_k(t), v_k(t)]^\mathrm{T}$ is the eigenstate of the dynamical B-dG equation. 
The dynamical Bd-G equation can be mapped to a Landau-Zener (LZ) problem~\cite{Zener1932, Damski2005, Dziarmaga2005, Cincio2007}.
Thus, we have the excitation probability of each $k$-mode as~\cite{Zurek2005}
\begin{eqnarray}\label{lz}
p_k=e^{-\frac{2\pi \tau_Qk^2}{\hslash}}.
\end{eqnarray}
The topological density of the final state is the summation of $k$, i.e., $n_d=\sum_k p_k$.
In the thermodynamic limit $N\rightarrow \infty$, $n_d$ can be obtained by a Gaussian integral as
\begin{eqnarray}
n_d=\frac{1}{2\pi}\int_{-\pi}^{\pi}p_k\mathrm{d}k=\frac{1}{2\pi}\sqrt{\frac{\hslash}{2J\tau_Q}}.
\end{eqnarray}
This result reveals that the topological defect density has a power-law scaling with respect to the quench rate~\cite{Zurek2005}
\begin{eqnarray}
n_d\sim \tau_Q^{-1/2},
\end{eqnarray}
which conforms to the prediction of KZM.

\begin{figure}
\begin{center}

\includegraphics[width=0.45\textwidth]{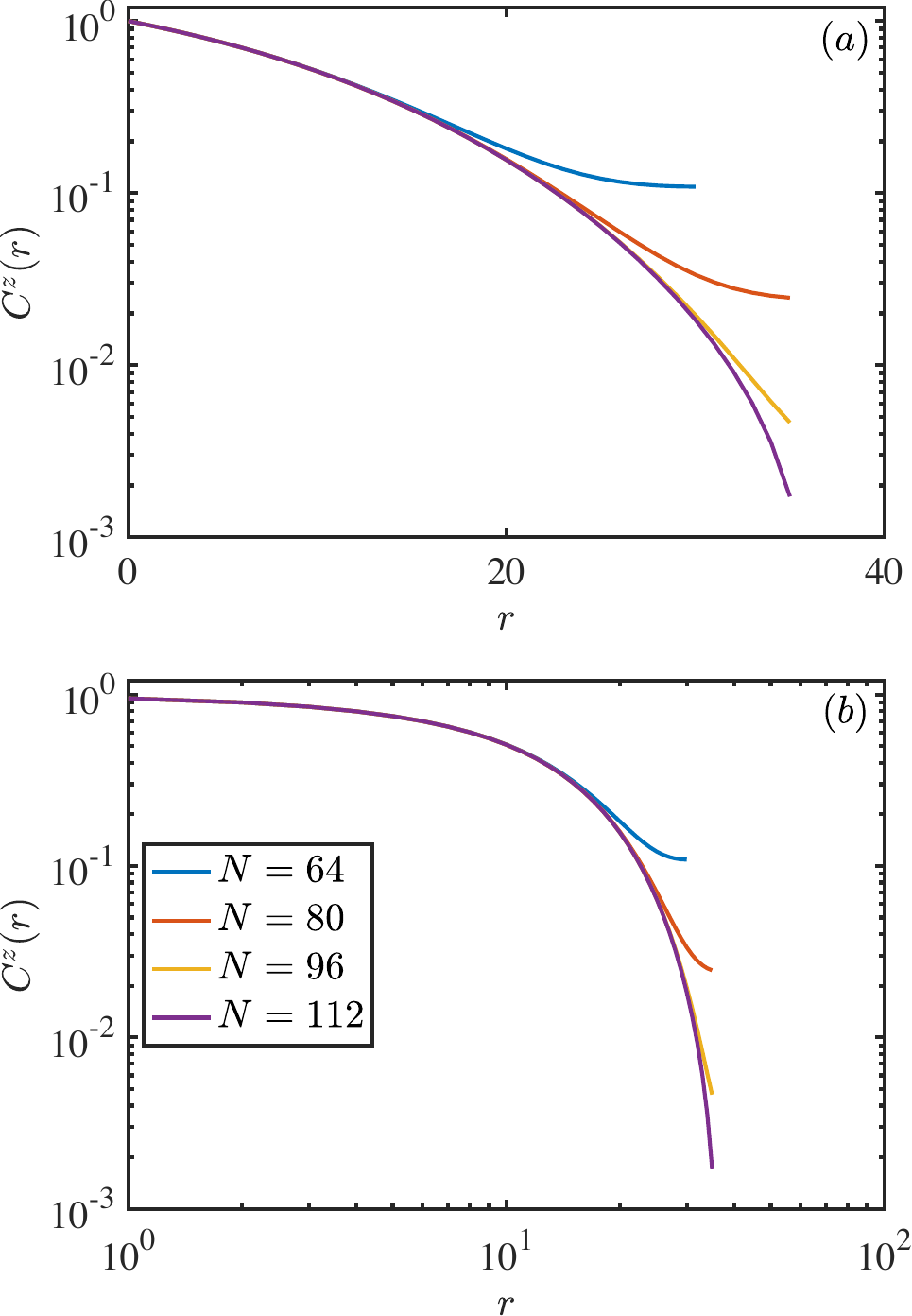}
\caption{\label{fig1}
Correlation functions of the final state after a linear quench for different sizes with (a) linear-logarithmic axis and (b) logarithmic-logarithmic axis.
Here, the corresponding quench rate is $\tau_Q=20$.
To avoid the boundary effect, we fix one operator at the central position of the lattice, i.e., $C^z(r)=\left<\hat{\sigma}_{L/2}^z\hat{\sigma}_{L/2+r}^z\right>$.
}
\end{center}
\end{figure}

\subsection{Quantum Fisher information after a slow quantum quench}\label{quench}
Now we start to investigate the multipartite entanglement of the quantum Ising model after a slow quantum quench, i.e., considering the dynamics of following time-dependent Hamiltonian
\begin{eqnarray}\label{ht}
H(t)=-\sum_{n=1}\hat{\sigma}_n^z\hat{\sigma}_{n+1}^z -g(t)\sum_{n=1}\hat{\sigma}_n^x.
\end{eqnarray}
We begin with a ground state of large $g$ in the paramagnetic phase, and mainly consider the relation between the QFI density and the quench rate.
Generally, QFI depends on the corresponding generator $\hat O$, so we first need to choose a proper operator in Eq.~(\ref{qfi}).
For $\hat H(t)$, there is a global $\mathbb{Z}_2$ spin-flip symmetry, 
of which the transformation operator has the form
\begin{eqnarray}
	\hat P=\prod_{n=1}^N\hat{\sigma}_n^x.
\end{eqnarray}
Here, the initial state is a ground state of the paramagnetic phase, which is $\mathbb{Z}_2$ symmetry unbroken.
In addition, we consider an isolated system, so the final state after the quench dynamics is also symmetry unbroken.
For the final Hamiltonian, i.e., $g=0$, the corresponding symmetry unbroken ground state is a GHZ state
\begin{eqnarray}
	\frac1{\sqrt2}\left(\ket{\uparrow}^{\otimes N}+\ket{\downarrow}^{\otimes N}\right),
\end{eqnarray}
which has the maximum QFI density $N$ with respect to the operator $ \hat{O}=\sum_{n=1}^{N}\hat{\sigma}^z_n$.
Assuming that the quench is slow enough, we expect that the final state will be very close to the GHZ state.
Based on these considerations, we can choose the operator as $ \hat{O}=\sum_{n=1}^{N}\hat{\sigma}^z_n$ to calculate QFI.

According to Eq.~(\ref{qfi}) and $\braket{\hat{\sigma}_n^z}=0$ (due to the existence of spin-flip symmetry),
we can calculate the QFI density  as
\begin{eqnarray}\label{fq}
f_Q=\frac1N \left<\left(\sum_{n=1}^N\hat{\sigma}_n^z\right)^2\right>=\frac1N\sum_{m=1}^N\sum_{n=1}^N\left<\hat{\sigma}_m^z\hat{\sigma}_n^z\right>,
\end{eqnarray}
where $\braket{\cdot}$ means the expectation value with respect to the final state.
Now, one can find that $f_Q$ is the summation of all two-site spin-spin correlation functions.
In the thermodynamic limit, i.e., $N\rightarrow \infty$, $f_Q$ of the final state is
\begin{eqnarray}
f_Q=1+\sum_{r=1}^{N} C_z(r),
\end{eqnarray}
where $C_z(r)=\left<\hat{\sigma}_i^z\hat{\sigma}_{i+r}^z\right>$ is the correlation function with distance $r$.

\subsection{Linear Quench}
Here, we adopt the MPS-based method to study the slow quench dynamics with open boundary conditions.
We first apply the DMRG algorithm to obtain the ground state for $g=5$ as the initial state.
Then, utilizing the time-evolving block decimation~(TEBD) algorithm~\cite{Haegeman2017, Ehlers2017, ZaunerStauber2018, Ren2020},
we can calculate the wave function of the final state.
Thus, according to Eq.~(\ref{fq}), we can obtain the QFI density of the final state.
For the DMRG calculation, the maximum bond dimension is 100 with the truncation error smaller than $10^{-12}$.
For the TEBD calculation, we choose the second-order Suzuki-Trotter decomposition.
Furthermore, we also enlarge the maximum bond dimension and decrease the time of a single step to increase the precision of results,
where the truncation error is smaller than $10^{-11}$, and the minimum time step of the Suzuki-Trotter decomposition is $\delta t = 0.02$.

\begin{figure}
\begin{center}
	\includegraphics[width=0.45\textwidth]{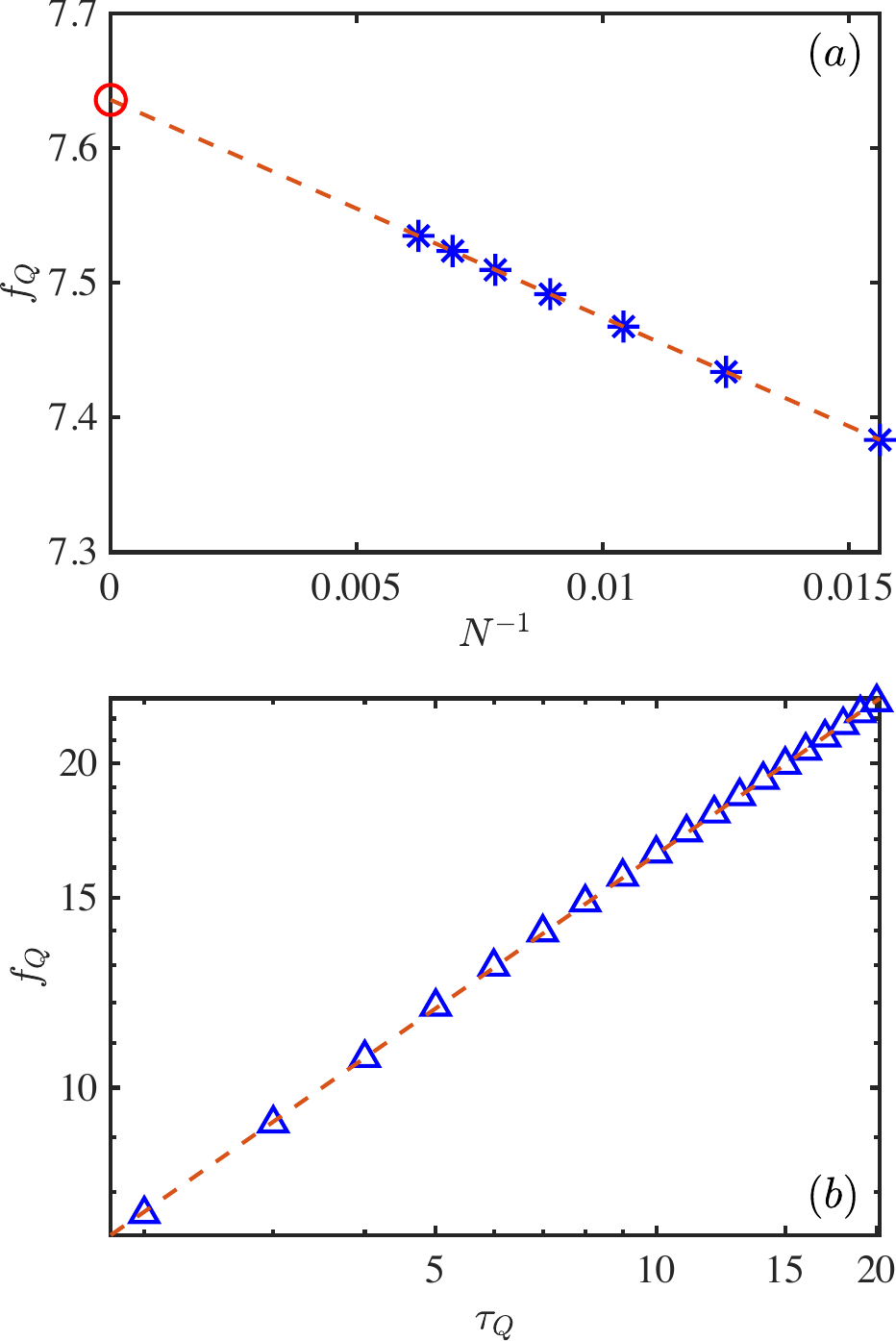}
	\caption{\label{fig2}
		The QFI density $f_Q$ after linear quenches crossing a quantum critical point.
		(a) Finite-size scaling of the QFI density of the final state $\tau_Q=2$.
		The scatters are the results of fixed finite size calculated by the MPS-based method, while the dashed line is the fitting curve.
		It indicates that $f_Q \approx f_Q^\infty-A/N$, where the intercept $f_Q^\infty$ (red circle) can be considered as the QFI density in the thermodynamic limit, and $A$ is a constant.
		Here, we obtain the interception value of about 7.6358.
		(b) The relation between the QFI density  $f_Q$ and the quench rate $\tau_Q$ in the thermodynamic limit.
		Each scatter is obtained by the method shown in (a).
		The dashed line is the linear fitting curve.
		  It is shown that $f_Q$ exhibits a power-law scaling with respect to $\tau_Q$, where the fitting result is $f_Q\sim\tau_Q^{0.474\pm0.002}\sim\tau_Q^{1/2}$.
	}
\end{center}
\end{figure}
Following the progress of the conventional KZM in the quantum Ising model, we first consider a linear slow quench.
The time-dependent transverse field has the form
\begin{eqnarray}
g(t)=-\frac t{\tau_Q} +1, \ \ \ \  -4\tau_Q\leq t \leq \tau_Q,
\end{eqnarray}
where $\tau_Q$ represents the quench rate and $t_c=0$.
The whole quench dynamics begins with $g=5$ and ends with $g=0$.

In Fig.~\ref{fig1}, we present the correlation function $C_z(r)$ of the final states.
It is shown that $C_z(r)$ decays very quickly as the increase of distance $r$, even faster than the exponential law~\cite{Cincio2007}.
Thus, the final states have no long-range correlation.
Now we start to consider the QFI density $f_Q$ of the final states.
To obtain more precise $f_Q$ in the thermodynamic limit, we first calculate $f_Q$ for a fixed $\tau_Q$ with different system sizes,
and then we use finite-size scaling to fit out $f_Q$ for $N\rightarrow0$, see Fig.~\ref{fig2}(a).
Finally, we study the relation between $f_Q$ and quench rate $\tau_Q$.
From Fig.~\ref{fig2}(b), we can find that 
\begin{eqnarray}\label{fQ}
f_Q\sim\tau_Q^{0.474\pm0.002}.
\end{eqnarray}
Therefore, the number of entangled spins after a slow quench has a power-law scaling with respect to the quench rate.

As mentioned in Sec.~\ref{kzm}, the conventional KZM predicts that the topological defect density of the final state satisfying the scaling $n_d\sim\tau_Q^{-d\nu/(z\nu+1)}$.
For the quantum Ising model, the topological defect is the domain wall $n_d(i):= \frac{1}{2}(1-\left<\hat{\sigma}_i^z\hat{\sigma}_{i+1}^z\right>)$ and $d=\nu=z=1$,
so $n_d\sim\tau_Q^{-\frac{1}{2}}$.
Therefore, after a linear quench, the QFI density $f_Q$ has a nearly inverse scaling to the one of the topological defects with a small correction.
We note that this small correction should not be the error of the numerical simulations.
The accuracy of our numerical simulations can be improved by enlarging the bond dimension and decreasing the time step of the Suzuki-Trotter decomposition,
and the final result in Eq.~(\ref{fQ}) has been converged and is accurate enough.

Now we try to understand the above results.
For the QFI density, according to the behavior of the correlation function $C_z(r)$, we can find the main contributions to $f_Q$ originate from $C_z(r<\xi)$,
where $\xi$ can be regarded as the correlation length of the final state.
That is, the QFI density is closely related to the correlation length.
Generally, after a linear quench crossing critical point, the correlation length is the inverse of topological defect density, i.e., $\xi\sim\tau_Q^{\nu/(z\nu+1)}$~\cite{Cincio2007}.
Therefore, the scaling of the QFI density may approximately satisfy
\begin{eqnarray}\label{fq_sca}
f_Q \sim\tau_Q^{d\nu/(z\nu+1)}.
\end{eqnarray}
However, according to the results in Fig.~\ref{fig1}, 
the correlation function $C_z(r)$ does not simply exhibit an exponential decay.
From Refs.~\cite{Cherng2006, Cincio2007}, we know that the correlation length of the final states in the quantum Ising model is complex,
where there exist corrections in the vicinity of $\sqrt{\tau_Q}$.
This correction of the correlation length can hardly affect the short-range physics.
However, it can result in an unnegligible correction for the nonlocal observables, e.g., QFI.
Therefore, QFI can reflect the nontrivial long-range physics after a slow quench dynamics crossing a critical point,
which is beyond the conventional KZM.

\subsection{Nonlinear Quench}
\begin{figure}	
\begin{center}
	\includegraphics[width=0.45\textwidth]{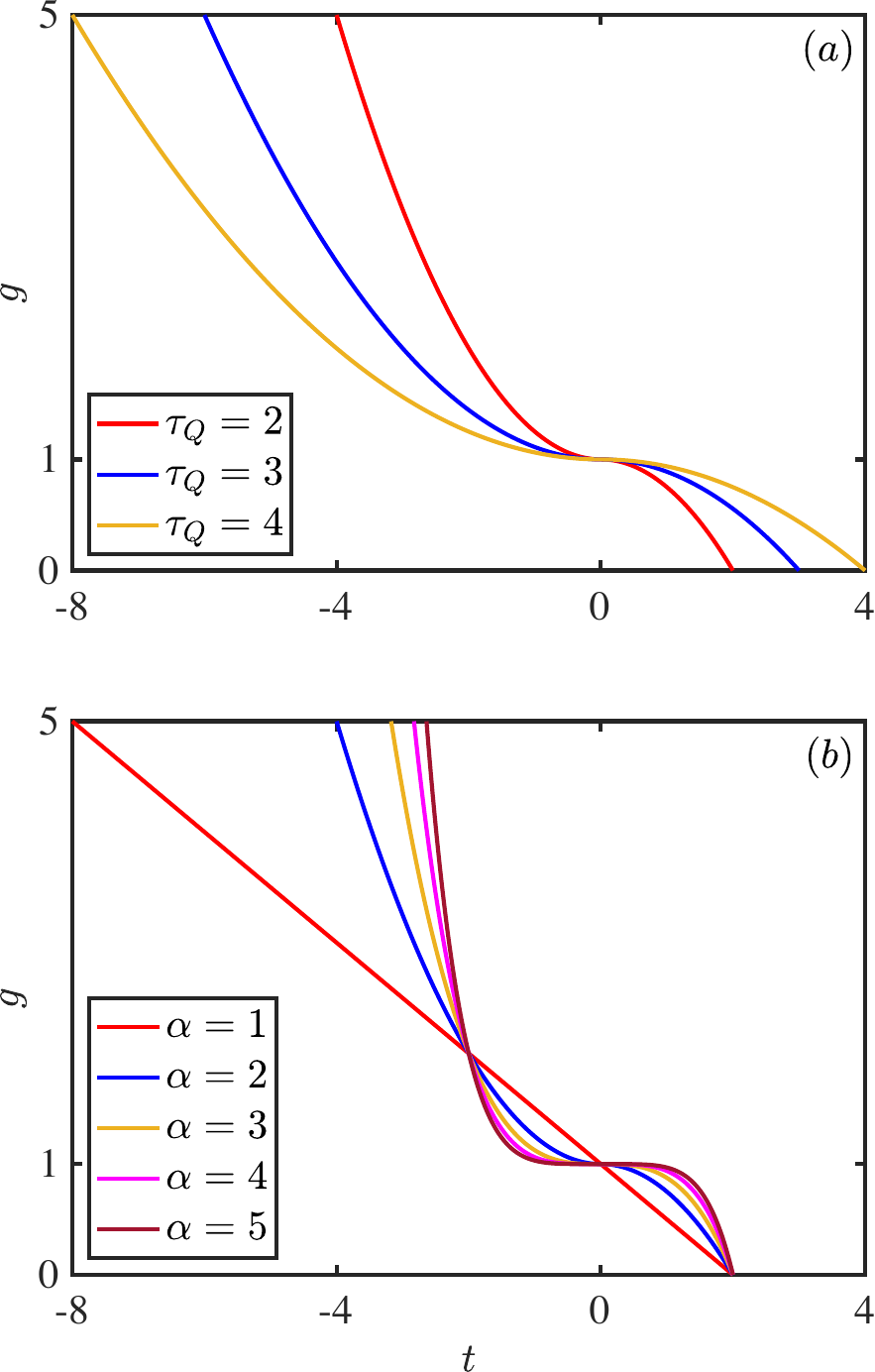}
	\caption{\label{fig3}
		Transverse field $g$ driven versus time $t$ for nonlinear quench.
		(a) Fixed $\alpha=2$ with different quench rates and (b) fixed $\tau_Q=2$ with different $\alpha$.
}
\end{center}
\end{figure}

To further illustrate the results in Eq.~(\ref{fq_sca}), 
we study the QFI density after a nonlinear quench.
Hence, we can use the following power-law function to study the general nonlinear quench
\begin{eqnarray}\label{gt}
g(t)=-\text{sgn}(t)\bigg|\frac{t}{\tau_Q} \bigg|^\alpha+1, 
\end{eqnarray}
where $\alpha>1$, see Fig.~\ref{fig3}.
In addition, similar to the linear case, the nonlinear quench dynamics begins with $g=5$ and ends with $g=0$.

\begin{figure}	
\begin{center}
	\includegraphics[width=0.45\textwidth]{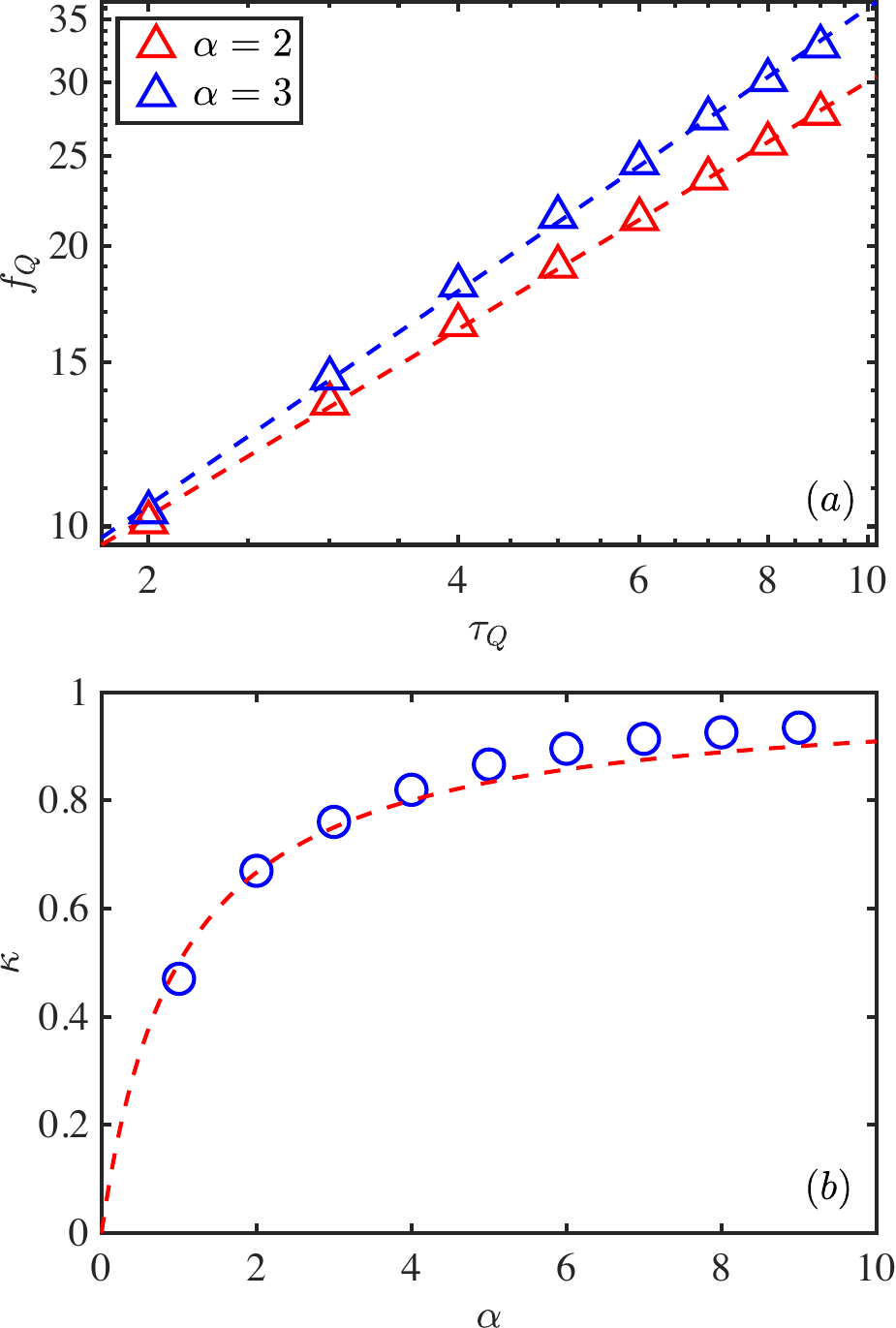}
	\caption{
		The QFI density $f_Q^{\text {nl}}$ after nonlinear quenches crossing a quantum critical point.
		(a) The relations between the QFI density $f_Q^{\text {nl}}$ and the quench rate $\tau_Q$ in the thermodynamic limit for different $\alpha$.
		Each scatter is obtained by the finite-size scaling in Fig.~\ref{fig2}(a), and the dashed line is the linear fitting curve.
		The fitting result is $f_Q^{\text {nl}}\sim\tau_Q^{0.66\pm0.02}\sim\tau^{2/3}_Q$ for $\alpha=2$ 
		and $f_Q^{\text {nl}}\sim\tau_Q^{0.76\pm0.02}\sim\tau^{3/4}_Q$ for $\alpha=3$.
		(b) The relation between scaling exponent of $f_Q^{\text {nl}}$ (labeled by $\kappa$) and $\alpha$.
		Each circle is the fitted results of the power-law scaling exponent for different $\alpha$.
		The red dash curve is the function image of $\kappa(\alpha)=\frac\alpha{\alpha+1}$.
	}
	\label{fig4}
\end{center}
\end{figure}

Here, as mentioned in Sec.~\ref{kzm}, the nonlinear quench of Eq.~(\ref{gt}) can be considered as a linear quench with respect to the driven parameter $\lambda = (g-1)^{1/\alpha}$ with the critical exponents
\begin{eqnarray}\label{expon}
z_\lambda = z = 1, \ \ \ \ \nu_\lambda = \alpha\nu=\alpha.
\end{eqnarray}
Therefore, according to Eqs.~(\ref{fq_sca}, \ref{expon}),
the QFI density  of the final states after a nonlinear quench, labeled $f_Q^{\text {nl}}$,  should approximate
\begin{eqnarray}\label{fq_n}
	f_Q^{\text {nl}} \sim\tau_Q^{d\nu_\lambda/(z_\alpha\nu_\lambda+1)} = \tau_Q^{\alpha/(\alpha+1)}.
\end{eqnarray}

In Fig.~\ref{fig4}, we present the numerical results of the QFI density after nonlinear slow quenches. 
According to Fig.~\ref{fig4}(a), we can see that for $\alpha=2$, 
\begin{eqnarray}
f_Q^{\text {nl}} \sim \tau_Q^{0.66\pm0.02},
\end{eqnarray}
and for $\alpha=3$, 
\begin{eqnarray}
f_Q^{\text {nl}} \sim \tau_Q^{0.76\pm0.02}.
\end{eqnarray}

Here, we can find that $f_Q^{\text {nl}}$ indeed exhibits a power-law scaling with respect to $\tau_Q$, 
and the exponents approximate $\alpha/(\alpha+1)$. 
We also present the corresponding power-law exponents versus $\alpha$ in Fig.~\ref{fig4}(b), 
which are overall consistent with Eq.~(\ref{fq_n}). 
However, we can also find that the corresponding exponent is slightly smaller than $\alpha/(\alpha+1)$ for small $\alpha$,
while it is slightly larger than $\alpha/(\alpha+1)$ for large $\alpha$.
The above results demonstrate that QFI after a nonlinear slow quench across a QPT can also be predicted by KZM with small corrections.

\section{Lipkin-Meshkov-Glick model}\label{lmg}
To further demonstrate our results, we also investigate the scaling behaviors in the LMG model, 
which is a full-connected model. 
The time-dependent Hamiltonian reads
\begin{eqnarray}
H(t)=-\frac2N\sum_{m<n}\hat{\sigma}^z_m \hat{\sigma}^z_n - g(t)\sum_{n}\hat{\sigma}^x_n,
\end{eqnarray}
where $N$ is the number of spins, and $g(t)$ is an external field. 
The LMG model has a second-order phase transition at the critical point $g=g_c=1$. 
When $g<1$, the system is in the ferromagnetic phase, where the parity symmetry is spontaneously breaking. 
When $g>1$, the system is in a symmetry-unbreaking phase with non-degenerate ground states.

Introducing the total spin operator 
\begin{eqnarray}
\hat{S}_{x/z}=\sum_n \hat{\sigma}_n^{x/z},  
\end{eqnarray}
we can rewrite the Hamiltonian as 
\begin{eqnarray}\label{angular}
H(t)=-\frac1N \hat{S}_z^2 - g(t) \hat{S}_x, 
\end{eqnarray}
which is convenient for numerical simulations.

The LMG model has no spatial dimensions. 
Compared with the quantum Ising model, the scaling behaviors of QFI after a slow quench may be different. 
In the LMG model, the corresponding parity operator associated to QPT reads
\begin{eqnarray}
\hat{P}=\prod_{n=1}^N \hat{\sigma}^x_n, 
\end{eqnarray}
which is the same as the quantum Ising model. 
When $g=0$, the final ground state is also a GHZ state, which has the maximum QFI density $N$, with respect to the operator $\hat{\sigma}_z$. 
Thus, we can also choose $\hat{\sigma}_z$ as the operator to calculate the QFI density. 

The LMG model is full-connected, so the QFI density of the final state scales as$f_Q \approx N$, i.e., almost all spins are entangled. 
According to the results of the quantum Ising model, we can find that crossing a critical point can decrease the multipartite entanglement.
Thus, for the LMG model, we can consider how many spins that are not entangled.
Here, we introduce the complement of the QFI density $f_Q$ to the maximum entanglement $N$, which is defined as 
\begin{eqnarray}
r_Q:=N-f_Q. 
\end{eqnarray}
In the following discussion, we mainly focus on the scaling of $r_Q$ with respect to the quench rate.

\begin{figure}	
	\begin{center}
		\includegraphics[width=0.6\textwidth]{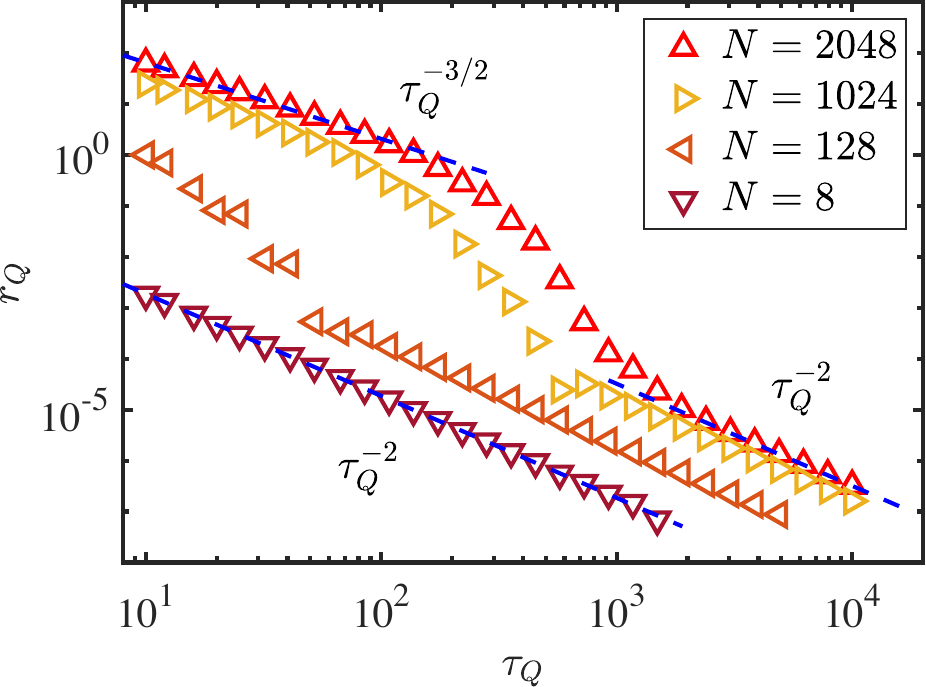}
		\caption{
			The complement of the QFI density $r_Q$ scales with respect to the quench rate $\tau_Q$. 
			Four groups of scatters correspond to system sizes $N=2048, 1024, 128, 8$ from top to bottom. 
			(1). The red scatters are of relatively large $N=2048$, and two scaling behaviors are presented distinctively. 
			Two blue dashed lines are the linear curve fitting in the logarithmic figure. 
			The left dash line exhibits a power-law fitting result $r_Q\sim\tau_Q^{-1.496\pm0.037}\sim\tau_Q^{-3/2}$, 
			and the right dashed line shows another power law $r_Q\sim\tau_Q^{-2.003\pm0.005}\sim\tau_Q^{-2}$. 
			(2). The bottom dashed line is the fitting result of $N=8$, where $r_Q \sim \tau_Q^{-2.002\pm0.015}$.
			\label{fig5}	}
	\end{center}
\end{figure}

For convenience, we only discuss the results of the linear quench. 
The transverse field satisfies 
\begin{eqnarray}
g(t)=-\frac t{\tau_Q} +1, \ \ \ \  -4\tau_Q\leq t \leq \tau_Q.
\end{eqnarray}
We drive the external field from $g=5$ to $0$. 
Here, the number of spins $N$ varies from $2^3$ to $2^{11}$, and $\tau_Q$ varies from $10^1$ to $10^5$.  
{Our numerical simulations are based on the Hamiltonian Eq.~(\ref{angular}), which is a $(N+1)\times(N+1)$ matrix. 
We obtain the ground state at $t=-4\tau_Q$ by the exact diagonalization as the initial state. 
The time evolution is obtained by solving an ordinary differential equation. 
}

In Fig.~\ref{fig5}, we show the numerical results of $r_Q$ versus $\tau_Q$, where there are two scaling behaviors. 
For instance, when $N=2^{11}$, for small $\tau_Q$, $r_Q$ scales as 
\begin{eqnarray}\label{rq_num}
r_Q\sim\tau_Q^{-1.496\pm0.037}.
\end{eqnarray} 
For large $\tau_Q$, $r_Q$ scales as  
\begin{eqnarray}
r_Q\sim\tau_Q^{-2.003\pm0.005}.
\end{eqnarray}

Now, we try to understand the two scaling behaviors. 
The complement $r_Q$ can be considered as the residual energy~\cite{Caneva2008}, defined as 
\begin{eqnarray}
E_\text{res}=E_\text{final}-E_\text{gs}, 
\end{eqnarray} 
where $E_\text{final}$ is the energy of the final state, i.e., the expectation value of $\hat{H}(\tau_Q)$ with respect to the final state, 
and $E_\text{gs}$ is the ground state energy of $\hat{H}(\tau_Q)$. 
The final Hamiltonian is 
\begin{eqnarray}
\hat{H}(\tau_Q)=-\frac1N \hat{S}_z^2. 
\end{eqnarray}
When $g=0$, due to the symmetry breaking, all spins are polarized in the same direction. 
Thus, the ground state energy $E_\text{gs}=-N$. 
Here, the residual energy can be written as 
\begin{eqnarray}
E_\text{res}=N-\frac 1N \left<\left(\hat{S}_z\right)^2\right>=N-f_Q=r_Q, 
\end{eqnarray}
which can also be regarded as the excitation energy. 
In the LMG model,  at the critical point $g_c=1$, the energy gap decays with the system size as 
\begin{eqnarray}\label{gaps}
\Delta\sim N^{-1/3},
\end{eqnarray}  
i.e., the dynamical exponent is $z=1/3$~\cite{Caneva2008, Acevedo2014}. 
For small $\tau_Q$, we can use the LZ formula~\cite{Zener1932, Damski2005} to estimate the scaling of $r_Q$, see Eq.~(\ref{lz}). 
The LZ formula gives the probability that a system driven through an energy gap is 
\begin{eqnarray}\label{p_lz}
p\sim e^{-\frac{\pi\Delta^2\tau_Q}{2\hslash}}.
\end{eqnarray} 
According to Eq.~(\ref{gaps}, \ref{p_lz}), we have $r_Q$, also the residual energy scales as 
\begin{eqnarray} \label{rq_kzm}
r_Q \sim \left(\frac{2\hslash|\ln p|}{\pi\tau_Q}\right)^{-3/2}\sim \tau_Q^{-3/2}, 
\end{eqnarray} 
which is consistent with the scaling of small $\tau_Q$~\cite{Zurek2005, Dziarmaga2010}. 
For a larger $\tau_Q$, i.e., $\tau_Q>\Delta$, we know that the LZ formula becomes invalid,
so the scaling of QFI cannot conform to the predictions of KZM~\cite{Caneva2008, Acevedo2014}.
We note that, for a local system, if  $\tau_Q>\Delta$, the adiabatic condition recovers,
where the topological defect density will exhibit an exponential decay with respect to $\tau_Q$.

According to Eqs.~(\ref{rq_num}, \ref{rq_kzm}), we can find that the scaling of QFI in the LMG model can be well predicted by KZM,
where there are hardly corrections.
Here, as a full-connected system, the LMG model has no spatial dimensions,
where the long-range physics cannot be defined.

\section{Discussion}\label{discussion}
Generally, near the critical point, the relaxation time of the system is divergent, where the divergent exponent depends on critical exponents.
This phenomenon is known as critical slowing down, which can be regarded as the origin of KZM.
Here, based on the above results, we try to understand the phenomenon of critical slowing down in QPTs from the viewpoint of multipartite entanglement.

For the quantum Ising model, due to the Lieb-Robinson bounds~\cite{Lieb1972, Bravyi2006}, the minimum time 
for creating a state with extensive multipartite entanglement from a short-range entangled state scales linearly in the size of the system.
That is, if the local system size is $N$, and the QFI density of the initial state is $O(1)$,
the time cost for creating a state with $O(N)$ QFI density is at least $O(N)$.
In the quantum Ising model, for the linear quench, if we want $f_Q\sim N$ for the final state, then the quench rate $\tau_Q\sim N^2$. 
Thus, we need time $t\sim N^2$ to create a state with the extended QFI density via a linear quench crossing a QPT, which is slower than the adiabatic case.
For the LMG model, we can also find that the crossing of the critical point can decrease the multipartite entanglement,
where the scaling of QFI is closely related to the critical exponents.

Therefore, compared with the adiabatic process, 
the rate of generating the multipartite entanglement near the critical point becomes slow
and is closely related to the critical exponents.
This behavior of the multipartite entanglement crossing a QPT can be considered as a reflection of critical slowing down.

\section{Summary and Outlook}\label{summary}
In summary, we have investigated the multipartite entanglement after slow quantum quenches crossing QPTs. 
For the quantum Ising model, 
by numerical calculations via MPS-based methods, 
we find that the QFI density of the final state follows the scaling behavior nearly opposite to the one of KZM with a small correction.
We demonstrate that this correction originates from the long-range physics after a slow quench.
We have also studied the LMG model, where the results conform better to KZM, since the long-range physics is absent in this full-connected system.
Our results reveal that KZM may not describe the long-range behaviors after a slow quantum quench crossing a critical point,
where the multipartite entanglement is a suitable observable to demonstrate these nontrivial long-range physics.
Moreover, our results can also enhance the understanding of QPTs and the behavior of critical slowing down from another viewpoint,
i.e., from the perspective of multipartite entanglement.
In addition, since QFI is accessible in quantum simulation experiments, we expect that our results can be realized experimentally in state-of-art quantum simulators.

There also remain many interesting open questions, which are deserved to be further studied.
In this work, we only consider 1D and full-connected systems, and it is meaningful to generalize our results to higher-dimensional systems. 
Moreover, when calculating the QFI density, we choose the corresponding local operators based on the symmetries of the Hamiltonians.
In these systems, the QPTs originate from spontaneous symmetry breaking, so we can use the order parameter to define QFI.
Therefore, it will be an interesting issue to study the multipartite entanglement after slow quantum quenches crossing a topological QPT,
which is absent of spontaneous symmetry breaking.
Furthermore, maybe other quantifications of multipartite entanglement could be adopted to testify to the generality of the scaling behavior.

\ack
We would like to thank Yu-Ran Zhang and Rui-Zhen Huang for helpful discussions. 
In this work, the numerical results are obtained by Tensor Network Python~(TeNPy)~\cite{Hauschild2018}, 
and qutip~\cite{Johansson2012, Johansson2013}. 
This work is supported by 
the National Natural Science Foundation of China (Grants Nos. 92265207, T2121001), 
and 
Beijing Natural Science Foundation (Grant No. Z200009).

\section*{References}
\bibliographystyle{unsrt}

\begin{thebibliography}{10}

\bibitem{Hyllus2012}
Philipp Hyllus, Wies{\l}aw Laskowski, Roland Krischek, Christian Schwemmer,
  Witlef Wieczorek, Harald Weinfurter, Luca Pezz{\'{e}}, and Augusto Smerzi.
\newblock Fisher information and multiparticle entanglement.
\newblock {\em Phys. Rev. A}, 85(2):022321, feb 2012.

\bibitem{Toth2012}
G{\'{e}}za T{\'{o}}th.
\newblock Multipartite entanglement and high-precision metrology.
\newblock {\em Phys. Rev. A}, 85(2):022322, feb 2012.

\bibitem{Hauke2016}
Philipp Hauke, Markus Heyl, Luca Tagliacozzo, and Peter Zoller.
\newblock Measuring multipartite entanglement through dynamic susceptibilities.
\newblock {\em Nat. Phys.}, 12(8):778--782, mar 2016.

\bibitem{Zhang2018}
Yu-Ran Zhang, Yu~Zeng, Heng Fan, J.{\hspace{0.167em}}Q. You, and Franco Nori.
\newblock Characterization of topological states via dual multipartite
  entanglement.
\newblock {\em Phys. Rev. Lett.}, 120(25):250501, jun 2018.

\bibitem{Pezze2017}
Luca Pezz{\`{e}}, Marco Gabbrielli, Luca Lepori, and Augusto Smerzi.
\newblock Multipartite entanglement in topological quantum phases.
\newblock {\em Phys. Rev. Lett.}, 119(25):250401, dec 2017.

\bibitem{Wu2005}
L.-A. Wu, S.~Bandyopadhyay, M.~S. Sarandy, and D.~A. Lidar.
\newblock Entanglement observables and witnesses for interacting quantum spin
  systems.
\newblock {\em Phys. Rev. A}, 72(3):032309, sep 2005.

\bibitem{Braunstein1994}
Samuel~L. Braunstein and Carlton~M. Caves.
\newblock Statistical distance and the geometry of quantum states.
\newblock {\em Phys. Rev. Lett.}, 72(22):3439--3443, may 1994.

\bibitem{Giovannetti2006}
Vittorio Giovannetti, Seth Lloyd, and Lorenzo Maccone.
\newblock Quantum metrology.
\newblock {\em Phys. Rev. Lett.}, 96(1):010401, jan 2006.

\bibitem{Pezze2016}
Luca Pezzè, Yan Li, Weidong Li, and Augusto Smerzi.
\newblock Witnessing entanglement without entanglement witness operators.
\newblock {\em Proc. Natl. Acad. Sci.}, 113(41):11459--11464, sep 2016.

\bibitem{Giovannetti2011}
Vittorio Giovannetti, Seth Lloyd, and Lorenzo Maccone.
\newblock Advances in quantum metrology.
\newblock {\em Nat. Photonics}, 5(4):222--229, mar 2011.

\bibitem{Peng2020}
Yi~Peng and Heng Fan.
\newblock Feedback ansatz for adaptive-feedback quantum metrology training with
  machine learning.
\newblock {\em Phys. Rev. A}, 101(2):022107, feb 2020.

\bibitem{Wang2021}
Zheng-An Wang, Yi~Peng, Dapeng Yu, and Heng Fan.
\newblock Beating standard quantum limit via two-axis magnetic susceptibility
  measurement.
\newblock {\em Chin. Phys. B}, 2021.

\bibitem{Amico2008}
Luigi Amico, Rosario Fazio, Andreas Osterloh, and Vlatko Vedral.
\newblock Entanglement in many-body systems.
\newblock {\em Rev. Mod. Phys.}, 80(2):517--576, may 2008.

\bibitem{Sachdev2009}
Subir Sachdev.
\newblock {\em Quantum Phase Transitions}.
\newblock Cambridge University Press, 2009.

\bibitem{Suzuki2013}
Sei Suzuki, Jun ichi Inoue, and Bikas~K. Chakrabarti.
\newblock {\em Quantum Ising Phases and Transitions in Transverse Ising
  Models}.
\newblock Springer Berlin Heidelberg, 2013.

\bibitem{Zeng2019}
Bei Zeng, Xie Chen, Duan-Lu Zhou, and Xiao-Gang Wen.
\newblock {\em Quantum Information Meets Quantum Matter}.
\newblock Springer New York, 2019.

\bibitem{Fromholz2020}
P.~Fromholz, G.~Magnifico, V.~Vitale, T.~Mendes-Santos, and M.~Dalmonte.
\newblock Entanglement topological invariants for one-dimensional topological
  superconductors.
\newblock {\em Phys. Rev. B}, 101:085136, Feb 2020.

\bibitem{Xia2009}
Y.~Xia, D.~Qian, D.~Hsieh, L.~Wray, A.~Pal, H.~Lin, A.~Bansil, D.~Grauer, Y.~S.
  Hor, R.~J. Cava, and M.~Z. Hasan.
\newblock Observation of a large-gap topological-insulator class with a single
  dirac cone on the surface.
\newblock {\em Nat. Phys.}, 5(6):398--402, may 2009.

\bibitem{Xu2018}
Kai Xu, Jin-Jun Chen, Yu~Zeng, Yu-Ran Zhang, Chao Song, Wuxin Liu, Qiujiang
  Guo, Pengfei Zhang, Da~Xu, Hui Deng, Keqiang Huang, H.~Wang, Xiaobo Zhu,
  Dongning Zheng, and Heng Fan.
\newblock Emulating many-body localization with a superconducting quantum
  processor.
\newblock {\em Phys. Rev. Lett.}, 120(5):050507, feb 2018.

\bibitem{Kitaev2006}
Alexei Kitaev and John Preskill.
\newblock Topological entanglement entropy.
\newblock {\em Phys. Rev. Lett.}, 96(11):110404, mar 2006.

\bibitem{Levin2006}
Michael Levin and Xiao-Gang Wen.
\newblock Detecting topological order in a ground state wave function.
\newblock {\em Phys. Rev. Lett.}, 96(11):110405, mar 2006.

\bibitem{Li2008}
Hui Li and F.~D.~M. Haldane.
\newblock Entanglement spectrum as a generalization of entanglement entropy:
  Identification of topological order in non-abelian fractional quantum hall
  effect states.
\newblock {\em Phys. Rev. Lett.}, 101(1):010504, jul 2008.

\bibitem{Eisert2010}
J.~Eisert, M.~Cramer, and M.~B. Plenio.
\newblock Colloquium: Area laws for the entanglement entropy.
\newblock {\em Rev. Mod. Phys.}, 82(1):277--306, feb 2010.

\bibitem{Vidal2003}
G.~Vidal, J.~I. Latorre, E.~Rico, and A.~Kitaev.
\newblock Entanglement in quantum critical phenomena.
\newblock {\em Phys. Rev. Lett.}, 90:227902, Jun 2003.

\bibitem{Holzhey1994}
Christoph Holzhey, Finn Larsen, and Frank Wilczek.
\newblock Geometric and renormalized entropy in conformal field theory.
\newblock {\em Nucl. Phys. B}, 424(3):443--467, aug 1994.

\bibitem{Cardy2016}
John Cardy.
\newblock Quantum quenches to a critical point in one dimension: some further
  results.
\newblock {\em J. Stat. Mech: Theory Exp.}, 2016(2):023103, feb 2016.

\bibitem{Osterloh2002}
A.~Osterloh, Luigi Amico, G.~Falci, and Rosario Fazio.
\newblock Scaling of entanglement close to a quantum phase transition.
\newblock {\em Nature}, 416(6881):608--610, apr 2002.

\bibitem{Guehne2005}
Otfried Gühne, G{\'{e}}za T{\'{o}}th, and Hans~J Briegel.
\newblock Multipartite entanglement in spin chains.
\newblock {\em New J. Phys.}, 7:229--229, nov 2005.

\bibitem{Hofmann2014}
Martin Hofmann, Andreas Osterloh, and Otfried Gühne.
\newblock Scaling of genuine multiparticle entanglement close to a quantum
  phase transition.
\newblock {\em Phys. Rev. B}, 89(13):134101, apr 2014.

\bibitem{Movassagh2016}
Ramis Movassagh and Peter~W. Shor.
\newblock Supercritical entanglement in local systems: Counterexample to the
  area law for quantum matter.
\newblock {\em Proc. Natl. Acad. Sci.}, 113(47):13278--13282, nov 2016.

\bibitem{Kibble1976}
T~W~B Kibble.
\newblock Topology of cosmic domains and strings.
\newblock {\em J. Phys. A: Math. Gen.}, 9(8):1387--1398, aug 1976.

\bibitem{Kibble1980}
T.W.B. Kibble.
\newblock Some implications of a cosmological phase transition.
\newblock {\em Phys. Rep.}, 67(1):183--199, dec 1980.

\bibitem{Zurek1985}
W.~H. Zurek.
\newblock Cosmological experiments in superfluid helium?
\newblock {\em Nature}, 317(6037):505--508, oct 1985.

\bibitem{Zurek1996}
W.H. Zurek.
\newblock Cosmological experiments in condensed matter systems.
\newblock {\em Phys. Rep.}, 276(4):177--221, nov 1996.

\bibitem{Laguna1997}
Pablo Laguna and Wojciech~Hubert Zurek.
\newblock Density of kinks after a quench: When symmetry breaks, how big are
  the pieces?
\newblock {\em Phys. Rev. Lett.}, 78(13):2519--2522, mar 1997.

\bibitem{Yates1998}
Andrew Yates and Wojciech~H. Zurek.
\newblock Vortex formation in two dimensions: When symmetry breaks, how big are
  the pieces?
\newblock {\em Phys. Rev. Lett.}, 80(25):5477--5480, jun 1998.

\bibitem{Dziarmaga1999}
Jacek Dziarmaga, Pablo Laguna, and Wojciech~H. Zurek.
\newblock Symmetry breaking with a slant: Topological defects after an
  inhomogeneous quench.
\newblock {\em Phys. Rev. Lett.}, 82(24):4749--4752, jun 1999.

\bibitem{Polkovnikov2005}
Anatoli Polkovnikov.
\newblock Universal adiabatic dynamics in the vicinity of a quantum critical
  point.
\newblock {\em Phys. Rev. B}, 72(16):161201, oct 2005.

\bibitem{Zurek2005}
Wojciech~H. Zurek, Uwe Dorner, and Peter Zoller.
\newblock Dynamics of a quantum phase transition.
\newblock {\em Phys. Rev. Lett.}, 95(10):105701, sep 2005.

\bibitem{Dziarmaga2005}
Jacek Dziarmaga.
\newblock Dynamics of a quantum phase transition: Exact solution of the quantum
  ising model.
\newblock {\em Phys. Rev. Lett.}, 95(24):245701, dec 2005.

\bibitem{GomezRuiz2019}
F.~J. G\'omez-Ruiz and A.~del Campo.
\newblock Universal dynamics of inhomogeneous quantum phase transitions:
  Suppressing defect formation.
\newblock {\em Phys. Rev. Lett.}, 122:080604, 2019.

\bibitem{GomezRuiz2020}
Fernando~J. G\'omez-Ruiz, Jack~J. Mayo, and Adolfo del Campo.
\newblock Full counting statistics of topological defects after crossing a
  phase transition.
\newblock {\em Phys. Rev. Lett.}, 124:240602, Jun 2020.

\bibitem{Lee2015}
Minchul Lee, Seungju Han, and Mahn-Soo Choi.
\newblock Kibble-zurek mechanism in a topological phase transition.
\newblock {\em Physical Review B}, 92(3):035117, July 2015.

\bibitem{Liou2018}
Shiuan-Fan Liou and Kun Yang.
\newblock Quench dynamics across topological quantum phase transitions.
\newblock {\em Physical Review B}, 97(23):235144, June 2018.

\bibitem{Roosz2014}
Gerg\ifmmode \mbox{\H{o}}\else~\H{o}\fi{} Ro\'osz, Uma Divakaran, Heiko Rieger,
  and Ferenc Igl\'oi.
\newblock Nonequilibrium quantum relaxation across a
  localization-delocalization transition.
\newblock {\em Phys. Rev. B}, 90:184202, Nov 2014.

\bibitem{Huang2020}
Rui-Zhen Huang and Shuai Yin.
\newblock Kibble-zurek mechanism for a one-dimensional incarnation of a
  deconfined quantum critical point.
\newblock {\em Phys. Rev. Research}, 2(2):023175, may 2020.

\bibitem{Barankov2008}
Roman Barankov and Anatoli Polkovnikov.
\newblock Optimal nonlinear passage through a quantum critical point.
\newblock {\em Phys. Rev. Lett.}, 101(7):076801, aug 2008.

\bibitem{Bando2020}
Yuki Bando, Yuki Susa, Hiroki Oshiyama, Naokazu Shibata, Masayuki Ohzeki,
  Fernando~Javier G\'omez-Ruiz, Daniel~A. Lidar, Sei Suzuki, Adolfo del Campo,
  and Hidetoshi Nishimori.
\newblock Probing the universality of topological defect formation in a quantum
  annealer: Kibble-zurek mechanism and beyond.
\newblock {\em Phys. Rev. Research}, 2:033369, Sep 2020.

\bibitem{Kou2023}
Han-Chuan Kou and Peng Li.
\newblock Varying quench dynamics in the transverse ising chain: The
  kibble-zurek, saturated, and presaturated regimes.
\newblock {\em Physical Review B}, 108(21):214307, December 2023.

\bibitem{Suzuki2024}
Fumika Suzuki and Wojciech~H. Zurek.
\newblock Topological defect formation in a phase transition with tunable
  order.
\newblock {\em Physical Review Letters}, 132(24):241601, June 2024.

\bibitem{Jamadagni2024}
Amit Jamadagni, Javad Kazemi, and Arpan Bhattacharyya.
\newblock Kibble-zurek mechanism and errors of gapped quantum phases.
\newblock {\em Physical Review B}, 110(4):045140, July 2024.

\bibitem{Chuang1991}
Isaac Chuang, Ruth Durrer, Neil Turok, and Bernard Yurke.
\newblock Cosmology in the laboratory: Defect dynamics in liquid crystals.
\newblock {\em Science}, 251(4999):1336--1342, mar 1991.

\bibitem{Du2023}
Kai Du, Xiaochen Fang, Choongjae Won, Chandan De, Fei-Ting Huang, Wenqian Xu,
  Hoydoo You, Fernando~J. Gómez-Ruiz, Adolfo del Campo, and Sang-Wook Cheong.
\newblock Kibble–zurek mechanism of ising domains.
\newblock {\em Nature Physics}, 19(10):1495--1501, June 2023.

\bibitem{Navon2015}
Nir Navon, Alexander~L. Gaunt, Robert~P. Smith, and Zoran Hadzibabic.
\newblock Critical dynamics of spontaneous symmetry breaking in a homogeneous
  bose gas.
\newblock {\em Science}, 347(6218):167--170, jan 2015.

\bibitem{Ko2019}
Bumsuk Ko, Jee~Woo Park, and Y.~Shin.
\newblock Kibble{\textendash}zurek universality in a strongly interacting fermi
  superfluid.
\newblock {\em Nat. Phys.}, 15(12):1227--1231, sep 2019.

\bibitem{Keesling2019}
Alexander Keesling, Ahmed Omran, Harry Levine, Hannes Bernien, Hannes Pichler,
  Soonwon Choi, Rhine Samajdar, Sylvain Schwartz, Pietro Silvi, Subir Sachdev,
  Peter Zoller, Manuel Endres, Markus Greiner, Vladan Vuleti{\'{c}}, and
  Mikhail~D. Lukin.
\newblock Quantum kibble{\textendash}zurek mechanism and critical dynamics on a
  programmable rydberg simulator.
\newblock {\em Nature}, 568(7751):207--211, apr 2019.

\bibitem{Monaco2006}
R.~Monaco, J.~Mygind, M.~Aaroe, R.~J. Rivers, and V.~P. Koshelets.
\newblock Zurek-kibble mechanism for the spontaneous vortex formation
  {inNb}-al/alox/{NbJosephson} tunnel junctions: New theory and experiment.
\newblock {\em Phys. Rev. Lett.}, 96(18):180604, may 2006.

\bibitem{Cui2020}
Jin-Ming Cui, Fernando~Javier G{\'{o}}mez-Ruiz, Yun-Feng Huang, Chuan-Feng Li,
  Guang-Can Guo, and Adolfo del Campo.
\newblock Experimentally testing quantum critical dynamics beyond the
  kibble{\textendash}zurek mechanism.
\newblock {\em Commun. Phys.}, 3(1):44, mar 2020.

\bibitem{Dziarmaga2006}
Jacek Dziarmaga.
\newblock Dynamics of a quantum phase transition in the random ising model:
  Logarithmic dependence of the defect density on the transition rate.
\newblock {\em Phys. Rev. B}, 74:064416, Aug 2006.

\bibitem{Polkovnikov2011}
Anatoli Polkovnikov, Krishnendu Sengupta, Alessandro Silva, and Mukund
  Vengalattore.
\newblock Colloquium: Nonequilibrium dynamics of closed interacting quantum
  systems.
\newblock {\em Rev. Mod. Phys.}, 83(3):863--883, aug 2011.

\bibitem{Grandi2010}
C.~De Grandi, V.~Gritsev, and A.~Polkovnikov.
\newblock Quench dynamics near a quantum critical point.
\newblock {\em Phys. Rev. B}, 81(1):012303, Jan 2010.

\bibitem{Polkovnikov2011a}
Anatoli Polkovnikov.
\newblock Microscopic diagonal entropy and its connection to basic
  thermodynamic relations.
\newblock {\em Ann. Phys.}, 326(2):486--499, feb 2011.

\bibitem{Mukherjee2008}
Victor Mukherjee, Amit Dutta, and Diptiman Sen.
\newblock Defect generation in a spin-$\frac{1}{2}$ transverse $xy$ chain under
  repeated quenching of the transverse field.
\newblock {\em Phys. Rev. B}, 77:214427, Jun 2008.

\bibitem{Grandi2010a}
C.~De~Grandi, V.~Gritsev, and A.~Polkovnikov.
\newblock Quench dynamics near a quantum critical point: Application to the
  sine-gordon model.
\newblock {\em Phys. Rev. B}, 81:224301, Jun 2010.

\bibitem{Cincio2007}
Lukasz Cincio, Jacek Dziarmaga, Marek~M. Rams, and Wojciech~H. Zurek.
\newblock Entropy of entanglement and correlations induced by a quench:
  Dynamics of a quantum phase transition in the quantum ising model.
\newblock {\em Phys. Rev. A}, 75(5):052321, may 2007.

\bibitem{Pollmann2010}
Frank Pollmann, Subroto Mukerjee, Andrew~G. Green, and Joel~E. Moore.
\newblock Dynamics after a sweep through a quantum critical point.
\newblock {\em Phys. Rev. E}, 81:020101, Feb 2010.

\bibitem{Dziarmaga2010}
Jacek Dziarmaga.
\newblock Dynamics of a quantum phase transition and relaxation to a steady
  state.
\newblock {\em Adv. Phys.}, 59(6):1063--1189, sep 2010.

\bibitem{Sen2008}
Diptiman Sen, K.~Sengupta, and Shreyoshi Mondal.
\newblock Defect production in nonlinear quench across a quantum critical
  point.
\newblock {\em Phys. Rev. Lett.}, 101(1):016806, jul 2008.

\bibitem{Zener1932}
Clarence Zener and Ralph~Howard Fowler.
\newblock Non-adiabatic crossing of energy levels.
\newblock {\em Proceedings of the Royal Society of London. Series A, Containing
  Papers of a Mathematical and Physical Character}, 137(833):696--702, 1932.

\bibitem{Damski2005}
Bogdan Damski.
\newblock The simplest quantum model supporting the kibble-zurek mechanism of
  topological defect production: Landau-zener transitions from a new
  perspective.
\newblock {\em Phys. Rev. Lett.}, 95(3):035701, jul 2005.

\bibitem{Haegeman2017}
Jutho Haegeman and Frank Verstraete.
\newblock Diagonalizing transfer matrices and matrix product operators: A
  medley of exact and computational methods.
\newblock {\em Annu. Rev. Condens. Matter Phys.}, 8(1):355--406, mar 2017.

\bibitem{Ehlers2017}
G.~Ehlers, S.~R. White, and R.~M. Noack.
\newblock Hybrid-space density matrix renormalization group study of the doped
  two-dimensional hubbard model.
\newblock {\em Phys. Rev. B}, 95(12):125125, mar 2017.

\bibitem{ZaunerStauber2018}
V.~Zauner-Stauber, L.~Vanderstraeten, M.~T. Fishman, F.~Verstraete, and
  J.~Haegeman.
\newblock Variational optimization algorithms for uniform matrix product
  states.
\newblock {\em Phys. Rev. B}, 97(4):045145, jan 2018.

\bibitem{Ren2020}
Jiajun Ren, Weitang Li, Tong Jiang, and Zhigang Shuai.
\newblock A general automatic method for optimal construction of matrix product
  operators using bipartite graph theory.
\newblock {\em J. Chem. Phys.}, 153(8):084118, aug 2020.

\bibitem{Cherng2006}
R.~W. Cherng and L.~S. Levitov.
\newblock Entropy and correlation functions of a driven quantum spin chain.
\newblock {\em Physical Review A}, 73(4):043614, April 2006.

\bibitem{Caneva2008}
Tommaso Caneva, Rosario Fazio, and Giuseppe~E. Santoro.
\newblock Adiabatic quantum dynamics of the lipkin-meshkov-glick model.
\newblock {\em Physical Review B}, 78(10):104426, September 2008.

\bibitem{Acevedo2014}
O.~L. Acevedo, L.~Quiroga, F.~J. Rodr\'{\i}guez, and N.~F. Johnson.
\newblock New dynamical scaling universality for quantum networks across
  adiabatic quantum phase transitions.
\newblock {\em Phys. Rev. Lett.}, 112:030403, Jan 2014.

\bibitem{Lieb1972}
Elliott~H. Lieb and Derek~W. Robinson.
\newblock The finite group velocity of quantum spin systems.
\newblock {\em Commun. Math. Phys.}, 28(3):251--257, sep 1972.

\bibitem{Bravyi2006}
S.~Bravyi, M.~B. Hastings, and F.~Verstraete.
\newblock Lieb-robinson bounds and the generation of correlations and
  topological quantum order.
\newblock {\em Phys. Rev. Lett.}, 97(5):050401, jul 2006.

\bibitem{Hauschild2018}
Johannes Hauschild and Frank Pollmann.
\newblock {Efficient numerical simulations with Tensor Networks: Tensor Network
  Python (TeNPy)}.
\newblock {\em SciPost Phys. Lect. Notes}, page~5, 2018.

\bibitem{Johansson2012}
J.R. Johansson, P.D. Nation, and Franco Nori.
\newblock Qutip: An open-source python framework for the dynamics of open
  quantum systems.
\newblock {\em Computer Physics Communications}, 183(8):1760--1772, August
  2012.

\bibitem{Johansson2013}
J.R. Johansson, P.D. Nation, and Franco Nori.
\newblock Qutip 2: A python framework for the dynamics of open quantum systems.
\newblock {\em Computer Physics Communications}, 184(4):1234--1240, April 2013.

\end{thebibliography}

\end{document}